\documentstyle[aaspp4]{article}

\def\sec{$^{\prime\prime}$} 
\def\deg{$^\circ$}

\slugcomment{Submitted to The Astrophysical Journal}

\lefthead{Storchi-Bergmann et al.}
\righthead{Optical continuum in Seyfert 2 galaxies}

\begin{document}

\title{THE NATURE OF THE OPTICAL LIGHT IN SEYFERT 2 GALAXIES WITH
POLARIZED CONTINUUM}

\author{Thaisa Storchi-Bergmann\altaffilmark{1}$^,$\altaffilmark{2}}
\affil{Departamento de Astronomia, IF-UFRGS,
	CP\,15051, CEP\,91501-970,
	Porto Alegre, RS, Brasil}
\authoremail{thaisa@if.ufrgs.br}

\author{Roberto Cid Fernandes\altaffilmark{2}}
\affil{Departamento de F\'\i sica, CFM-UFSC, Campus Universit\'ario Trindade,
       CP\,476, CEP\,88040-900,
       Florian\'opolis, SC, Brasil}

\author{Henrique R. Schmitt\altaffilmark{2}$^,$\altaffilmark{3}}
\affil{Departamento de Astronomia, IF-UFRGS,
	CP\,15051, CEP\,91501-970,
	Porto Alegre, RS, Brasil}

\altaffiltext{1}{Visiting Astronomer at the Cerro Tololo 
Interamerican Observatory, operated by the Association of Universities for 
Research in Astronomy, Inc. under contract with the 
National Science Foundation.}

\altaffiltext{2}{e-mail addresses: thaisa@if.ufrgs.br, cid@fsc.ufsc.br,
schmitt@if.ufrgs.br}

\altaffiltext{3}{CNPq Fellow}

\begin{abstract}

We investigate the nature of the optical continuum and stellar
population in the central kiloparsec of the Seyfert 2 galaxies
Mrk\,348, Mrk\,573, NGC\,1358 and Mrk\,1210 using high
signal-to-noise ratio long-slit spectra obtained along the radio
axis or along the extended high excitation emission.  These four
galaxies are known to have polarized continuum---including polarized
broad lines in Mrk\,348 and Mrk\,1210---and previous studies
indicate featureless continua contributions in the 20--50\% range at
$\lambda \approx$ 5500\AA. Nevertheless, our measurements of the
equivalent widths of absorption lines and continuum ratios as a
function of distance from the nuclei show no decrease of the
equivalent widths (i.e., no dilution) nor bluening of the spectrum
towards the nucleus, as expected if a blue featureless continuum was
present at the nucleus in the above proportions. 
We investigate one possibility to account for the lack of dilution:
that the stellar population at the nucleus is the same as that
from the surrounding bulge and dominates the nuclear light. By comparing the nuclear and the extranuclear spectra of each galaxy, we
conclude that this hypothesis works  for Mrk\,348, NGC\,1358 and
Mrk\,1210, for which we find stellar population contributions at the
nucleus larger than 90\% at all wavelengths. Our approach differs
from that adopted in previous studies, where an elliptical galaxy
template is used to represent the stellar population of the nucleus.
Although the latter  may be valid for some galaxies---as, for
example, Mrk\,573---in several cases the stellar population may
be different from that of an elliptical galaxy.

We find that a larger stellar population contribution to the nuclear
spectra can play the role of the ``second featureless continuum''
source (FC2) inferred from previous studies. In particular,
stellar population synthesis shows that the nuclear regions of
Mrk\,348 and Mrk\,1210 have important contributions of young to
intermediate age stars (0--$10^8yr$), not present
in templates of elliptical galaxies. In the case of
Mrk\,1210, this is further confirmed by the detection of a
``Wolf-Rayet feature'' in the nuclear emission-line spectrum.


\end{abstract}

\section{Introduction}

The development of the unified model for Seyfert galaxies has opened
interesting questions regarding the nature of the optical--UV
continuum in Seyfert 2s. This continuum seems to be well understood
only in the prototypical Seyfert 2 galaxy NGC\,1068 (Antonucci \&
Miller 1985, Miller, Goodrich \& Mathews 1991), where it is
attributed to a combination of a red stellar population contribution
plus a blue featureless continuum scattered towards us by electrons,
and in Mrk\,477 (Heckman et al. 1997), where the observed blue
continuum is dominated by a dusty starburst. Polarimetric studies by
Miller \& Goodrich (1990), and later by Tran, Miller \& Kay (1992),
Kay (1994) and Tran (1995a,b,c) have revealed polarized lines and
continua in a number of additional Seyfert 2's, indicating the
presence of a scattered Seyfert 1 component. However, the low levels
of polarization (Miller \& Goodrich 1990), lack of detectable broad
lines in the direct spectra (Cid Fernandes \& Terlevich
1995, Heckman et al. 1995), and larger polarizations in the broad
lines than in the continuum (Tran 1995a,b,c) indicate the existence
of another source of continuum, not accounted for by the starlight
template nor by the reflected nuclear component.  The origin of
``FC2'', as this extra component became known (Miller 1994, Tran
1995c) is one of the most important current issues in Seyfert 2s
studies. The two currently contending scenarios are free-free
emission from the scattering region (Tran 1995c), and young-stars,
as advocated by Cid Fernandes \& Terlevich (1995) and Heckman
et al. (1995).

When discussing the nature of the optical continuum in active
galaxies, one must be aware of the importance of evaluating and
removing the starlight ``contamination'', particularly in Seyfert
2s, where the stellar component often overwhelms the scattered flux.
Indeed, as emphasized by Miller \& Goodrich (1990), Antonucci
(1993), Kay (1994) and Tran (1995a), the derived polarizations are
very sensitive to the adopted stellar population template. The
simple fact that different authors obtain very different starlight
fractions for the same object (e.g., Kay 1994, Tran 1995a) is a
clear demonstration of how critical and uncertain is the starlight
removal procedure. In most polarimetric studies, as in fact in most
optical spectroscopy studies of AGN since Koski (1978), the stellar
population template used to subtract the stellar contribution is the
nuclear spectrum of an elliptical or early type galaxy. Given a
library of normal galaxy spectra spanning a sufficiently wide range of
stellar population mixtures, this approach will certainly produce a
satisfactorily AGN-looking, smooth residual spectrum. However, given
the non-uniqueness of the solution, one must always bear in mind
that the adopted template may not correspond to the true stellar
population of the host galaxy. And even if the adopted template
accounts for, say, 80\% of the true stellar component, the error
induced by not accounting for the remaining 20\% may be critical
when it comes to interpreting the nature of the residual continuum.

In order to investigate the stellar populations in active galaxies
using a different approach, we (Cid Fernandes, Storchi-Bergmann 
\& Schmitt 1997, hereafter Paper I) have recently concluded a
study where we have used long-slit optical spectroscopy of 42
galaxies to measure the variation of equivalent widths (hereafter W)
of several stellar absorption lines, as well as continuum ratios as
a function of distance from the nucleus.
Our goals were to map the stellar populations in the
sample galaxies, and to obtain the stellar population
characteristics right at the active nucleus via extrapolation of the
properties measured in neighboring extranuclear regions. Such an
approach has been previously used by Fosbury et al. (1978) and by
ourselves in previous works (e.g. Schmitt, Storchi-Bergmann \&
Baldwin 1994; Storchi-Bergmann et al. 1995, 1997a).

In Paper I we found that the presence of a featureless continuum in
the nucleus of Seyfert 1 galaxies was revealed by a bluening of the
continuum and by a decrease of the W's of the absorption lines
towards the nucleus. In galaxies with nuclear starbursts or rings of
star formation, the same behavior was observed:  the star-forming
regions present bluer colors and smaller W's. One unexpected result
was the finding that, in the majority of Seyfert 2 galaxies, there
is neither a dilution of the W's nor a bluening of the continuum
towards the nucleus, indicating that the featureless continuum, if
present at the nucleus, contributes with less than $\approx$10\% 
of the flux in the optical waveband.  The only clear 
cases of detection of dilutions
and blue continua were the Seyfert 2 galaxies already known to
harbor ``composite nuclei'', with both Seyfert and starburst activity,
as evidenced by the emission-line ratios and/or the presence of
absorption features of young stars. 
These results are illustrated
in Figure 1, where we show the variation of W(Ca II K), W(G band),
W(Mg I$+$MgH) and of the continuum ratio $\lambda$5870/$\lambda$4020 as
a function of distance from the nuclei for the Seyfert 1 galaxy 
NGC\,6814, for the galaxy with composite nucleus (Seyfert 2 and Starburst)
NGC\,7130 and the Seyfert 2 galaxy Mrk\,348. While the decrease
of the W's and bluening of the continuum is obvious in the two former
galaxies, they are not observed in the latter.

In this paper, we investigate the above results more closely for four
Seyfert 2 galaxies: Mrk\,348, Mrk\,573, NGC\,1358 and Mrk\,1210, as
previous works have reported dilution factors up to
$\sim$50\% for these galaxies.  In Paper I we have shown that our
method allows the detection of dilutions larger than 5--10\% 
(which correspond to the uncertainties in the W's),  
but for these galaxies we have not detected
significant dilutions in the absorption features of the stellar
population up to at least $\sim$ 10\sec\ from the nucleus. 
This behavior can be observed in Fig. 1 for Mrk\,348 and
in similar plots for  Mrk\,573, NGC\,1358 and Mrk\,1210 
(Figs. 17, 21 and 25  from Paper I). We
investigate the reason for this lack of dilution and the
implications for the nature of the nuclear continuum and stellar
population.

\section{The Data}

The data discussed here have been described in Paper I. In summary, 
they consist of long-slit spectra obtained with the Cassegrain 
Spectrograph and Reticon detector at the 4m telescope of the Cerro Tololo 
Inter-American Observatory. The pixel size was $\approx$ 1\sec\ and
the slit width 2\sec. For the four galaxies discussed here,  the observations
were obtained under photometric conditions with a
seeing of $\approx$ 1\sec. The slit was
oriented along the radio axis and/or along the extended emission,
but the observations were scheduled to make this angle coincide
with the paralactic angle, in order to avoid the effects of
differential refraction. The spectral resolution (FWHM of the
calibration lamp lines) is $\approx$ 7\AA.
A summary of the observations is presented
in Table 1. The spectra were reduced and flux calibrated 
using standard procedures in IRAF,
and one-dimensional spectra were extracted along the spatial direction
in windows of 2\sec$\times$2\sec. Prior to the analysis, the spectra
were corrected for the foreground Milky Way reddening, using the
empirical selective extinction function of Cardelli, Clayton \& Mathis (1989).
The corresponding E(B-V) values are listed in the last column of
Table 1 and were extracted from the NASA Extragalactic Database.

\section{Extranuclear spectra as population templates for the nucleus}

Mrk\,348, Mrk\,573, NGC\,1358 and Mrk\,1210 are nearby galaxies,
whose bulges have effective radii ranging from 6\sec\ to 36\sec,
as determined by near-infrared surface photometry (Alonso-Herrero,
Ward \& Kotilainen, 1996; Heisler, De Robertis \& Nadeau 1996;
Xanthopoulos 1996). Thus, the bulge is completely resolved in
these galaxies, and spectra extracted at less than 6\sec\ 
from the nucleus will still be sampling the bulge stellar population.

We believe that using a stellar population template from the 
bulge of the same galaxy  to represent the stellar population of the nucleus 
is in principle a more robust approach than adopting a
population template from other galaxies, as Paper I shows that the stellar population varies substantially from galaxy to galaxy.
The smooth radial variation of the W's of absorption lines found in Paper I, 
also supports this approach. The W's only show abrupt changes when
there is evidence of the presence of a blue stellar population---
in the starburst galaxies or in the rings of star formation
---or a featureless continuum---in the Seyfert 1 galaxies.

Moreover, as pointed out in the introduction, previous works usually adopt an
elliptical galaxy template to represent the stellar population 
in Seyfert nuclei, whereas recent studies (Wyse, Gilmore \& Franx 1997)
show that stellar populations in bulges frequently differ from
that in ellipticals. Jablonka, Martin \& Arimoto (1996) show that 
bulges occupy regions similar to that of ellipticals in the fundamental plane,
but are displaced towards smaller line strengths. Balcells \&
Peletier (1994) showed that, although bulges follow a color-magnitude
relationship similar to that of ellipticals, they have a larger scatter 
and are bluer than ellipticals of similar luminosity.
The use of an elliptical galaxy template to subtract the stellar population
contribution of the bulge of an spiral or lenticular galaxy 
is thus likely to produce a residual blue continuum due to a ``template mismatch'' (Cid Fernandes \& Terlevich 1992).

These facts, together with the discrepant results obtained by different 
authors using the traditional elliptical galaxy template method,
motivated us to seek for an alternative starlight evaluation technique.
In this section we compare the bulge extranuclear spectra 
with the nuclear spectra of
Mrk\,348, Mrk\,573, NGC\,1358 and Mrk\,1210,
in order to investigate if the former can 
be used as stellar population templates for the latter. 

Two criteria were used to select the 
windows for the extranuclear spectra: (1) to keep the smallest
possible angular distance from the nucleus---always smaller than the
effective radius of the bulge---in order to minimize changes in the
stellar population characteristics; (2) to
ensure that the distance was large enough to avoid
contamination from nuclear light  due to seeing effects.

The nuclear and extranuclear windows all have sizes of
2\sec$\times$2\sec. 
In order to construct the extranuclear spectra, we have skipped
the windows adjacent to the nucleus and  
averaged the two extractions centered at 4\sec\ at both sides of the
nucleus, which include light from 3\sec\ to 5\sec\ from the nucleus.
As the seeing was $\approx$1\sec\ in the observations, we do not expect significant contamination by nuclear light at these windows.
The  corresponding average distances from the nuclei
at the galaxies are 1.2, 1.4, 1.0 and 1.0 kpc,
respectively for Mrk\,348, Mrk\,573, NGC\,1358 and Mrk\,1210.

We compare the nuclear with the extranuclear optical (3500 to
6000\AA) spectrum of each galaxy in Figs. 2 to 5.
As all galaxies present extended
emission,  we have chopped the emission lines from the extranuclear
spectra in order to allow a more clear comparison of the absorption
features. An extranuclear spectrum is considered to be an acceptable fit
to the nuclear one if the nuclear W's are reproduced within 10\% and the
continuum within 5\%. At the blue end of the spectra, somewhat larger
differences in the continua were allowed ($\approx$10\%) due to
the larger noise in this spectral region. 
These values correspond to the maximum 
uncertainties in the measurements of W's and continua in Paper I.
The residuals in the continuum were calculated as 
averages of the difference spectra (nuclear$-$extranuclear),
 in windows avoiding emission lines. The difference spectra are
also shown in Figs. 2--5. 
 
In all cases, the nuclear continuum is slightly redder than the
extranuclear one (see Paper I). Assuming that this is due to an excess
reddening of the nuclear relative to the extranuclear spectra (in
agreement with the accepted scenario that the nuclear regions in
Seyfert 2 galaxies are dusty), we have reddened the extranuclear
spectra by the necessary
amount to make the continuum slopes to match that of the nuclear
spectra. We now discuss the results for each galaxy.

\subsection{Mrk\,348}

Figure 2a shows the comparison of the nuclear with the reddened
[E(B-V)=0.03] extranuclear spectrum of Mrk\,348, after normalization
of the latter to the flux of the nuclear spectrum near 5500\AA,
as well as the residual between the two. 
The nuclear continuum is well reproduced (within 5\%) by the
extranuclear one after a small reddening of the latter, and, in Paper
I, we have also concluded that the equivalent widths of absorption
lines are the same within the uncertainties ($\le$10\%). These results
indicate that the nuclear stellar population can be reproduced by that
of the extranuclear spectrum; no extra spectral component seems to be
needed within the margins of our matching criteria.

We now compare our results with those obtained by
Tran~(1995a,b,c). Tran has shown that Mrk\,348 presents a Seyfert
1 spectrum in polarized light, with a continuum polarization of
5.7\% (after starlight correction), increasing to more than twice
this value in broad H$\alpha$. The stellar component, represented by
the elliptical galaxy NGC 821, was found to contribute 73\% of the
total flux at $\lambda$5500, while our results suggest a much larger
fraction. Tran's analysis of the spectropolarimetry data also showed
that FC1 (the scattered Seyfert 1 component in Tran's nomenclature),
amounts to 5\% of the flux at $\lambda$5500, and that it has a
$F_\lambda \propto \lambda^{-1}$\ shape. Unlike the starlight
fraction, the FC1 strength is not strongly dependent on the choice
of template, since Tran's determination takes into account the
strength necessary to match the polarization of the continuum to
that of the broad lines. Similarly, the $\lambda^{-1}$\ shape is not
critically affected by the template, as it is implicit in the
polarized spectrum ($P_\lambda \times F_\lambda$). Since we know that
a scattered component is present, the spectral decomposition in
Fig.~2a is not physically acceptable, as it corresponds to 100\%
starlight. Yet, as shown below, there is room for a small
contribution from the scattered component.

Adopting a F$_\lambda\propto \lambda^{-1}$\ shape for the reflected
continuum, a contribution of 5\% at $\lambda$5500\AA\ corresponds to a
decrease of $\approx$5\% of W(MgI+MgH), somewhat larger for W(G band),
and up to $\approx$10\% for W(CaII K), thus within our measurement
uncertainties for the W's. In order to illustrate that both W's and continuum
shape are in agreement with the presence of such a scattered nuclear
continuum, we have combined the extranuclear spectrum with a
F$_\lambda\propto \lambda^{-1}$\ power-law in proportions of 95\% and
5\% at $\lambda$5500\AA, respectively, and compared this spectrum with
the nuclear one.  The best reproduction of the nuclear spectrum was
obtained when, prior to the combination with the power-law component,
we reddened the extranuclear spectrum by an additional
E(B-V)=0.06. The result  is shown in Figure 2b.

The essential difference between Tran's results and ours is that, while
he attributes the remaining 22\% of the photons at 5500 \AA\ to an
extra continuum source (FC2), we account for the whole spectrum with
either just the extranuclear stellar population template (Fig.~2a) or,
more realistically, with a combination of this template plus the
reflected nuclear spectrum (Fig.~2b). 

We remark that the slit width used in our
observations (2\sec), is similar to that used by Tran (2.4\sec), and
the same used by Kay (1994). The 2\sec\ length of our nuclear
extraction window was smaller that the typical 5--7\sec\ used by
Kay, while Tran does not give this information, but it is unlikely
that it is smaller than ours. We thus rule out the possibility that
we have found a larger galaxy contribution due to a larger nuclear
window. Instead, the different starlight fractions found by Tran and
us stem from the different stellar population templates used. Note
that Tran's galaxy fractions are already larger than the ones
obtained by Miller \& Goodrich (1990) for the objects in common,
including Mrk\,348 and Mrk\,1210. Our results show that an even
larger stellar contribution is possible.

\subsection{Mrk\,573}

Figure 3a shows the comparison of the nuclear spectrum of Mrk\,573
with the reddened [E(B-V)=0.1] extranuclear spectrum.  In this case,
we find some mismatch of the continuum between 4000 and 5000\AA, the
reddened extranuclear spectrum having $\ge$10\% less flux than the
nuclear spectrum in this spectral region.  Mrk\,573 has a polarized
continuum observed by Kay~(1994).  Allowing for a few percent
contribution of such component (represented by a F$_\lambda\propto
\lambda^{-1}$\ power-law, in analogy with Mrk\,348), we get an even
worse fit in the $\lambda\lambda$4000-5000\AA\ region.

We have then tried a different stellar population template, obtained
from the nuclear spectrum of the elliptical galaxy IC\,4889, from our
sample of Paper I (thus, observed with the same setup as Mrk\,573),
combined with the same power-law component above. Figure 3b shows that
this template, combined with a 10\% contribution of the power-law at
$\lambda$5500 gives a better fit for the continuum, and at the same
time reproduces the nuclear W values.  This result is in approximate
agreement with previous studies, like that of Kay (1994), who derived a
stellar population contribution of 80\% at $\lambda$4400\AA, while our
results imply a stellar population contribution of 85\% at this
wavelength.

But how to understand the lack of dilution of the W's in the nucleus
of this galaxy? One possibility is that the stellar population shows
a small variation with distance from the nucleus, more evident in
the shape of the continuum than in the absorption lines, and that
the scattered (power-law) component is {\it extended}, keeping the
W's approximately at the same values. An extended blue continuum,
probably due to scattered light, has indeed been found in near-UV
images of Mrk\,573 by Pogge \& De Robertis (1993) along the
ionization cones observed in this galaxy. If this interpretation is
correct, spatially resolved spectropolarimetry should reveal a
Seyfert 1 spectrum up to a few arcseconds from the nucleus along the cone
axis.

\subsection{NGC\,1358}

Figure 4a shows the comparison of the nuclear spectrum of NGC\,1358
with the reddened [E(B-V)=0.1] extranuclear spectrum.  The 
two spectra present $<$5\% differences in the continuum
and $<10$\% differences in the W's, suggesting no contribution by
a featureless continuum. Nevertheless, Kay (1994) finds a continuum
polarization of 1.65\% (after starlight subtraction), indicating the
presence of a scattered component, albeit weak. She derives a
stellar population contribution of 83\% at $\lambda4400$\AA, while
our results indicate a larger contribution, of at least 90\% at the
Ca II K line. Allowing for a contribution of up to 10\% of a
F$_\lambda\propto\lambda^{-1}$\ component in the blue, at
$\lambda$5500 this contribution amounts only to $\approx$3\%. The
fit we obtain for the nuclear spectrum, combining 3\% of this
component with 97\% of extranuclear spectrum at $\lambda$5500\AA\ is
slightly improved, as illustrated in Figure 4b. At $\lambda$4400 the
stellar population contribution  amounts to  $\approx$95\%. We note
that the 83\%  starlight fraction found by Kay yields an uncomfortably
low polarization for NGC 1358 as compared to the typical values
expected in the unified model (Miller \& Goodrich 1990). Only for
fractions above 95\%, and thus within the range of the 
acceptable stellar population contributions according to our method,
would the polarization raise above 10\%.

\subsection{Mrk\,1210}

Fig. 5a shows a comparison of the nuclear spectrum of Mrk\,1210
with the extranuclear one scaled and reddened by E(B-V)=0.07. 
The latter is a good reproduction of the nuclear
spectrum down to the Ca II K line,
with differences in the continuum $<$5\% and $<$10\% in the W's. But
bellow 3900\AA, there is an excess continuum in the nuclear spectrum.  
This continuum could be due to the Balmer recombination
continuum of the emitting gas, which is significant in this galaxy
because of its strong emission lines.  The
gaseous temperature, as derived from the [OIII]$\lambda$4363/5007
emission-line ratio is about 20,000\deg K. We have then used the
corresponding H$\beta$ recombination coefficient and HI continuum
emission coefficient from Osterbrock (1989), and the observed
H$\beta$\ flux (5.38$\times$10$^{-14}$ ergs cm$^{-2}$ s$^{-1}$) 
to derive the contribution
of the Balmer continuum to the nuclear spectrum.
This contribution amounts to $\approx$3\% of the total flux at
$\lambda$5500, $\approx$7\% at $\lambda$3700 and 30\% for $\lambda
<$3646\AA. We note that a Balmer continuum has also been detected in
Mrk\,477 (Heckman et.\ al.\ 1997, hereafter H97), a galaxy in many
respects similar to Mrk\,1210 (see below).

Mrk\,1210 was also observed by Tran (1995a,b,c), who found a
polarized continuum contribution of 6\% at $\lambda$5500\AA. As for
Mrk\,348, this value is derived by Tran as being the one necessary
to make the polarization in the continuum to agree with that of the
broad lines. Tran finds a 75\% stellar contribution, while the
remaining 19\% ``non-stellar'' continuum is attributed to FC2. The
spectral distribution of the scattered continuum is derived to be
F$_\lambda \propto \lambda^{-0.7}$.  We have then combined this
power-law contribution  normalized to 6\% at $\lambda$5500\AA, the
Balmer continuum described above (3\% at $\lambda$5500\AA) and the
extranuclear spectrum normalized to 91\% of the nuclear flux at
$\lambda$5500\AA\ after applying an additional reddening of E(B-V)=0.08.
The result is the spectral distribution illustrated in Fig. 5b, which is
an improved fit to the nuclear spectrum. Again, the different 
stellar fractions obtained by us and previous
works result from the different stellar population templates. As for
Mrk\,348, what Tran (1995c) calls FC2 is somehow included in our
stellar population template. 

A close look at the nuclear emission-line spectrum shows that this
galaxy seems to  present a ``Wolf-Rayet feature'', which can  be
observed in Fig. 6a.  We point out the similarity of the whole
complex to that found in Mrk\,477 by H97. We identify 5 narrow
emission-lines in this complex: [FeIII]$\lambda$4658\AA,
HeII$\lambda$4686\AA, [ArIV]$\lambda\lambda$4711,4740\AA\ and
[NeIV]$\lambda$4725\AA, to which we have fitted narrow gaussians,
constraining their FWHM to the average value of the other strong
emission lines in the spectrum (720 km s$^{-1}$). The  subtraction
of these lines leaves the residual shown in Figure 6b, centered
around $\lambda$4670\AA, with  flux 1.22$\times10^{-14}$ erg
cm$^{-2}$ s$^{-1}$ (and luminosity $3.7\times10^{39}$ erg s$^{-1}$).

Following the methodology of H97 we have compared our spectrum with
those from WR stars (Smith, Shara \& Moffat 1996; Conti \& Massey
1989), in order to estimate the type of WR stars producing the emission
lines. We reach a similar conclusion: that the residual of Fig. 6b can
be identified as broad emission from HeII$\lambda$4686\AA\ and
NIII$\lambda$4640\AA, and the WR stars in Mrk1210 can be classified as
of subtype WN6-7 or later.

In a starburst, the ratio of the flux of the Wolf-Rayet complex to
the H$\beta$ flux gives a measure of the relative importance of WR
and O stars (H97 and references therein).  In Mrk\,1210, this value
is $\approx$20\%, which is a lower limit, as the H$\beta$\ flux
includes contribution from ionization by the AGN. This value is
twice the value found for Mrk\,477 (H97) before correcting for the
contribution from the AGN, and indicates the presence at the nucleus
of Mrk\,1210 of a stellar burst with an age between 3 and 8 Myr
(H97, Meynet 1995).

\section{Population Synthesis for the Stellar Population Templates}

Since we have concluded that the difference between the starlight
fractions obtained here with respect to those in previous works
(particularly for Mrk 348 and Mrk 1210) is due to the different
stellar population templates, it is natural to ask what do our
templates have that ordinary elliptical galaxies don't?

In order to answer this question, we have performed a population
synthesis analysis using the base of stellar clusters of Bica (1988)
and the method described in Schmitt, Bica \& Pastoriza (1996). As
input data we have used the W's of Ca II K, CN, G-band and Mg I +
MgH absorption features, as well as the continuum fluxes at 3660,
4020, 4570 and 6630\AA, all normalized to the flux at 5870\AA. 

The results of the synthesis are summarized in Table 2. The stellar
population of Mrk\,573 is represented by the elliptical template from
IC\,4889, in the last line of the Table, while for the other galaxies
the results correspond to the extranuclear templates discussed in \S3. The
columns of the table show the percentage contribution of the age bins
corresponding to 10 Gyr (globular clusters), 1 Gyr, 100 Myr, 10 Myr and
HII regions, respectively, to the spectrum at $\lambda$5870\AA. We also
list, between parentheses, the corresponding fractions at $\lambda$4510\AA.

Inspection of the table shows that in NGC\,1358, as in the case of
the elliptical galaxy template, the population is composed almost only by
the 10 Gyr and 1 Gyr age bins. In Mrk\,348 and Mrk\,1210, however,
there are additional contributions from the younger age bins
($\le$100 Myr), amounting to 10\% and 23\%,
respectively. Since these populations are blue, the corresponding
fractions increase for shorter wavelengths, reaching 14\% and 33\%
at 4510\AA\ respectively (Table 2). 
We thus conclude that the difference between our extranuclear templates
and the more commonly used elliptical galaxy spectrum is the presence
of a substantial contribution from a 0--100 Myr old population.

A combination of a red bulge, power-law continuum, young and
intermediate age components has been also found in previous works on
the analysis of the optical continuum of active galaxies.  Tadhunter,
Dickinson \& Shaw (1996), for example, have concluded that an
intermediate age component was required in order to reproduce the
optical spectrum of the nearby radio galaxy 3C\,321, in addition to a
power-law and an old bulge component.  Using a similar method as used
here, Storchi-Bergmann et al. (1997b) have also found a mixture of old
bulge, power-law and ``aging starburst'' components in the
continuum spectrum of the inner region of Cen A. Cid Fernandes \&
Terlevich (1992) have also suggested that the contribution from young
and intermediate age stars in Seyfert 2s may have been sistematically
underestimated due to a ``template mismatch'' effect. Such a mismatch
results from the known difficulty in determining the contribution of
early type stars to the optical spectrum of AGN (see discussions in
Miller \& Goodrich 1990 and Kay 1994). Further studies suggesting the
presence of young to intermediate age stars in active galaxies are
the ones about the strength of the Ca II triplet around 8500\AA\ 
(Terlevich, Diaz \& Terlevich 1990), and the infrared work of
Oliva et al. (1995).

\section{Analysis using an elliptical galaxy template}

Had we used an elliptical galaxy as a template for Mrk\,348 and
Mrk\,1210, we would inevitably obtain larger featureless continuum (FC)
contributions. Experiments using the nuclear spectrum of the elliptical
galaxy IC 4889 confirm this (see Figures 7 and 8). 
In order to match the nuclear spectrum of Mrk 348, a combination
of 85\% contribution of the elliptical template IC4889 and a 
15\% $\lambda^{-1}$ power-law contribution at 
$\lambda$5500\AA\ would be required. For
Mrk 1210, the proportions would be 69\% of starlight and 28\% of a
$\lambda^{-0.7}$\ power-law (plus 3\% contribution of the Balmer
continuum), in better agreement with Tran's (1995a) results.
 
It is important to emphasize that the fits obtained with an
elliptical galaxy are as good as those obtained with the extranuclear
templates! However, these are by no means equivalent choices, since the
differences in the starlight fractions imply completely different
scenarios for the nature of the residual non-stellar continuum. Which
leaves us with the question: How is one to decide which template to use?

Whilst it is not possible to decide {\it a priori} which template
suits best the true stellar population in the nucleus, we believe
that a spectrum extracted from the bulge
of the same galaxy, should be, in most cases, a better representation
of the nuclear stellar population
than a spectrum from an external galaxy, whose star formation
history may differ substantially from that of the target galaxy. 
Furthermore, the much higher FC fractions obtained above with the
elliptical galaxy template would imply W and color radial gradients,
if the FC is concentrated in or close to the nucleus, which should be
detectable with our method, whereas we find none (Paper I). For example,
Fig. 1 shows that the W(Ca II K) of NGC\,6814
decreases by $\approx$40\%  at the nucleus 
due to the FC. In the case of Mrk\,348, for the 15\% FC 
contribution at $\lambda$5500\AA\ derived above with the nuclear stellar
population represented by the elliptical galaxy template, 
the FC contribution at the Ca II K line
is 33\%, while for Mrk\,1210 the 28\% contribution at $\lambda$5500\AA\ 
corresponds to 45\% at the Ca II K line. Such dilutions are comparable to
that observed in NGC\,6814 and should thus be readily observed.

The only possibility to have an elliptical galaxy template plus
a FC at the nucleus and no radial dilution of the absorption lines
is if the FC is {\it extended}
and the FC to galaxy ratio remains constant throughout the inner regions.
In this case our extranuclear spectra would not correspond to the pure
stellar population component, since they would be contaminated by the FC
(be it FC1, FC2 or both).
Testing this possibility will require further observational constraints. Direct imaging in the UV could in principle reveal an extended continuum,
as was observed in Mrk\,573 (Pogge \& De Robertis 1993). Polarization maps
would allow to test the presence of extended scattered light,
as well as long-slit spectra obtained at several position angles.

Regardeless of its extension, the FC in Mrk\,348 and Mrk\,1210 is clearly
dominated not by scattered light, but by FC2. Tran (1995c) finds that
83\% of the FC is due to FC2 in Mrk\,348 and 76\% in Mrk\,1210. The difference
between the extranuclear templates of Mrk\,348 and 1210 
and an elliptical galaxy should thus be essentially
due to FC2. According to the synthesis results, this difference
is due to young and intermediate age stars, which we therefore associate
with FC2. A conclusive detection of features from
such stars would corroborate this idea.
The WR feature in Mrk\,1210 confirms the presence of young stars. UV
spectroscopy could provide further evidence of this through the
detection of SiIV and CIV absorptions, as found in Mrk\,477
(H97). In the case of Mrk\,348 the evidence for young to
intermediate age stars relies not on a direct detection, but on our
population synthesis analysis. UV spectroscopy of this object will
be crucial to test this result.

UV spectropolarimetry would also be useful to test the contribution
of young to intermediate age stars. While the older stellar
components should contribute little to the UV emission, the young
stars should extend into the UV range along with the scattered AGN
component. Since the stellar flux is presumably unpolarized, we
expect that Mrk\,1210 (and possibly Mrk\,348) presents broad
UV lines (e.g. CIV$\lambda$1550 and MgII$\lambda$2800) with a {\it larger
polarization} than the adjacent continuum. In other words, the
``dilution by FC2 effect'' found by Tran (1995c) should extend into
the UV.

\section{Summary and Concluding Remarks}

We have investigated the nature of the optical continuum and stellar
population in the nuclear region of the Seyfert 2 galaxies Mrk\,348,
Mrk\,573, NGC\,1358 and Mrk\,1210 using long-slit spectroscopic data.

In a previous work (Paper I) we have found that none of these galaxies
present dilutions in the nuclear equivalent widths W's of the absorption
lines as compared with those of the extranuclear spectra from the bulge of the same galaxy, as it would be expected if a featureless continuum was
present at the nuclei in the proportions found in previous works (20\%--50\%).
We have tested one possibility to account for this lack of 
dilution: that the nuclear stellar population is the same as the
extranuclear one from the bulge, and the featureless 
continuum contribution to the nuclear spectra 
is smaller than 10\% (and thus not producing enough dilution in the W's
to be detectable with our method). 
We conclude that this hypothesis works for Mrk\,348, NGC\,1358 and Mrk\,1210.

A larger stellar population 
contribution can solve several puzzles of Seyfert 2 galaxies:
(1) it raises the level of intrinsic polarization, bringing it
closer to the values expected in the unified model (Miller \&
Goodrich 1990); (2) it avoids the problem of overpredicting the
strengths of broad lines in the direct spectra (Cid Fernandes \&
Terlevich 1992, 1995, Heckman et al. 1995); and (3) it allows a
compatibilization of the polarizations measured for the broad lines
and continua. This last point is implicit in our results, since we
have shown that, for Mrk 348 and Mrk 1210, the nuclear spectrum can
be adequately fit with a stellar component plus the scattered
continuum at the strengths inferred from spectropolarimetry (Tran
1995a,b,c). Since these are exactly the problems which lead to the idea of a
second featureless continuum component (Miller 1994), our results suggest 
that FC2 is the result of an underestimation of the stellar
contribution.

Our population synthesis analysis indicates that the larger stellar
population contribution found for Mrk\,348 and Mrk\,1210 relative to
that found in previous works is due to the presence of young and
intermediate age stars ($\le 100$~Myr), not accounted for by the more
commonly used elliptical galaxy templates. In other words,
the FC2 would be due to blue stars in these two cases. 
For Mrk\,1210, the apparent
detection of a ``Wolf-Rayet'' feature confirms this result. These findings,
together with results from the recent study by H97 on Mrk\,477, suggest
that young populations may be common in Markarian
Seyferts, being responsible for at least part of the blue continuum in
these galaxies (which were discovered precisely due to their blue
continuum).

The above results suggest that star-formation is an
important piece in the AGN puzzle. We speculate that, if there is 
a causal link between star-formation and activity, 
the $\sim 100$\ Myr star formation time scale
identified in this paper should be comparable to the
duration of the nuclear activity cycle.
Since 100 Myr is about 1\% of the 
Hubble time, a causal link between star formation and activity would
imply that about 1\% of the galaxies should be active.
Interestingly, this value seems to agree with current
estimates (e.g. Huchra \& Burg 1992). Population synthesis for
a complete sample of nearby Seyferts could help putting
tighter constraints on this link.

We have also tested an elliptical galaxy template,
as used in previous works. For Mrk\,573, this template 
gives a better match to the nuclear stellar population than the extranuclear spectrum. The lack of dilution in this case could be due to the presence of
an extended scattered continuum (observed in UV images 
by Pogge \& De Robertis, 1993). Mrk\,348 and Mrk\,1210 
also present an excess blue continuum when compared with 
an elliptical template. According to our W's measurements, this
continuum should be extended, but according to the spectropolarimetry
by Tran (1995c), only  a small part of it is polarized, so it is
unlikely that it is dominated by scattered light. 
This result is thus consistent with our finding of an excess
blue population in these galaxies relative to an elliptical galaxy.

As a final remark, we note that, as all current FC2 indicators
are critically dependent on the starlight correction procedure,
unveiling the nature of FC2 will necessarily require a carefull
examination of the stellar population in the central regions of AGN.
With this goal, in a forthcoming paper (Schmitt et al. 1998), 
we investigate the stellar content of a sample of 42 active and
normal galaxies, via population synthesis as a function of distance
from the nuclei, using the extensive data of Paper I.

\acknowledgments
We acknowledge fruitful
discussions with Laura Kay, Andrew S. Wilson and Roberto Terlevich,
and valuable suggestions by an anonymous referee. 
This research received partial support from the Brazilian
institutions CNPq, Finep, FAPERGS and FAPEU, and has made use of 
the NASA/IPAC Extragalactic Database (NED), which is operated by the Jet 
Propulsion Laboratory, Caltech, under contract with NASA.

%

%
%

\clearpage

\begin{figure}
\plotone{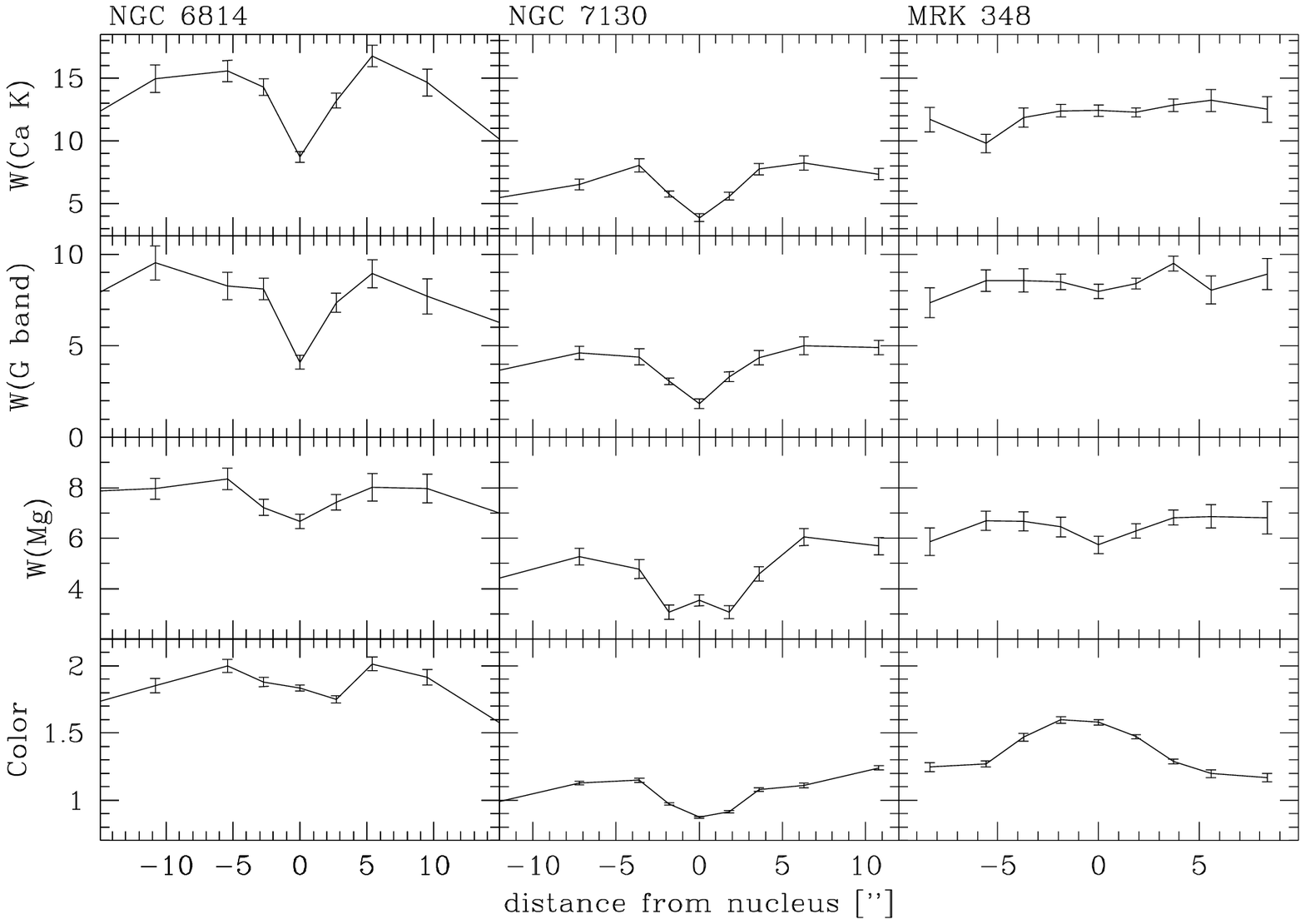}
\caption{From top to bottom: Radial variations of the equivalent 
widths of the absorption features Ca II K ($\lambda$3930\AA), 
G-band ($\lambda$4301\AA), Mg I + MgH ($\lambda$5176\AA) and
continuum ratio $\lambda$5870/$\lambda$4020 for the Seyfert 1
galaxy NGC\,6814 (left), the composite Seyfert $+$ Starburst
galaxy NGC\,7130 (center) and the Seyfert 2 Mrk\,348 (right). (Adapted
from Paper I).}
\end{figure}

\begin{figure}
\plotone{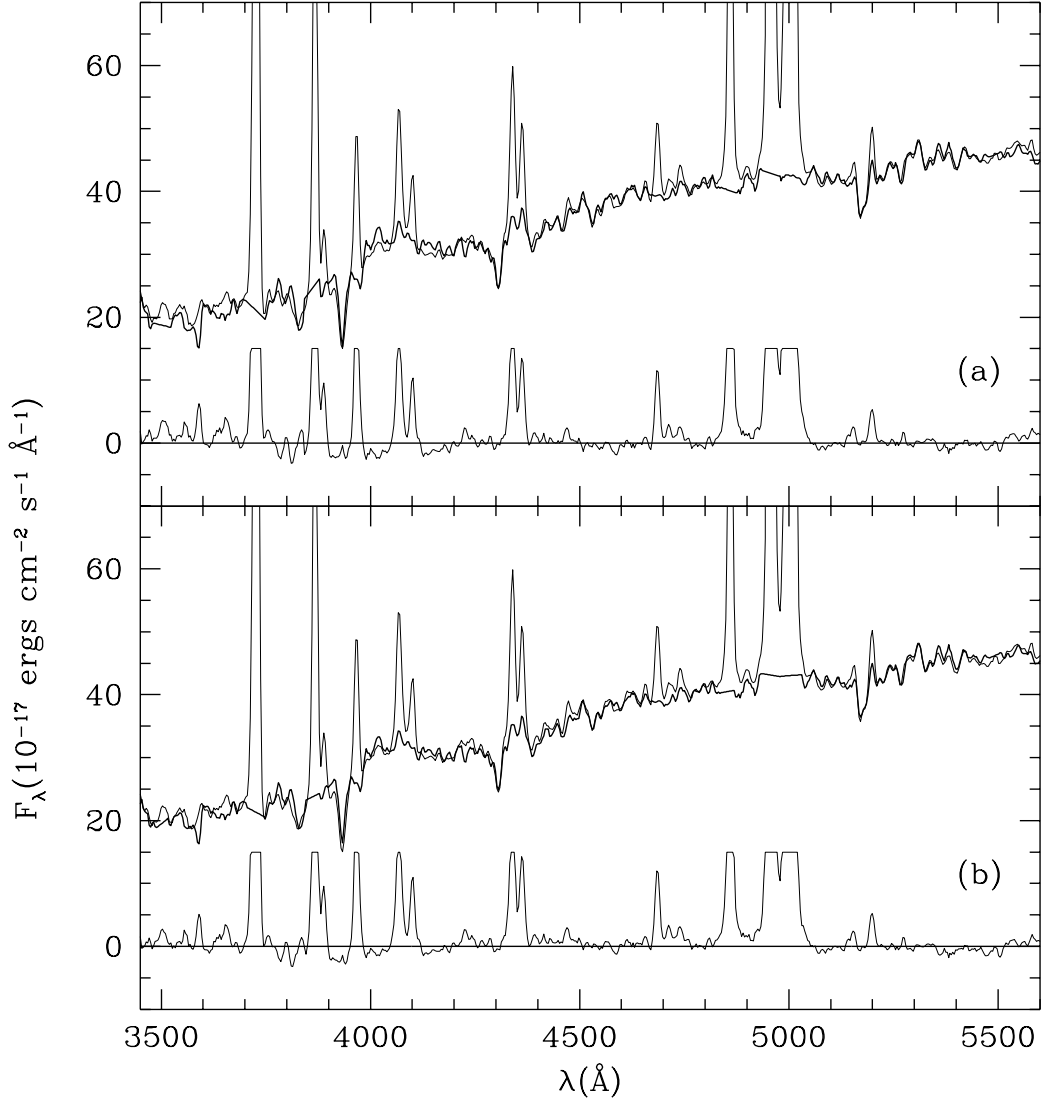}
\caption{Comparison between the nuclear and extranuclear bulge
spectra of Mrk\,348: (a) The nuclear spectrum  (thin line) is plotted
on top of the normalized and reddened [E(B-V)=0.03]  
extranuclear bulge spectrum (thick line), 
where the emission lines have been chopped for clarity.
The residual between the two is plotted at the bottom of the figure,
showing that the nuclear stellar population is well reproduced by
that of the bulge. 
(b) The nuclear spectrum (thin line) is compared with the
extranuclear spectrum reddened by E(B-V)=0.09, combined with
5\% contribution (at $\lambda$5500\AA) of the FC1 component,
represented by a $F_\lambda \propto \lambda^{-1}$ power-law (thick line). 
The residual between the two is plotted at the bottom, showing 
a slightly improved fit when compared with (a). The emission lines
in the residual spectra have also been chopped for clarity.}
\end{figure}

\begin{figure}
\plotone{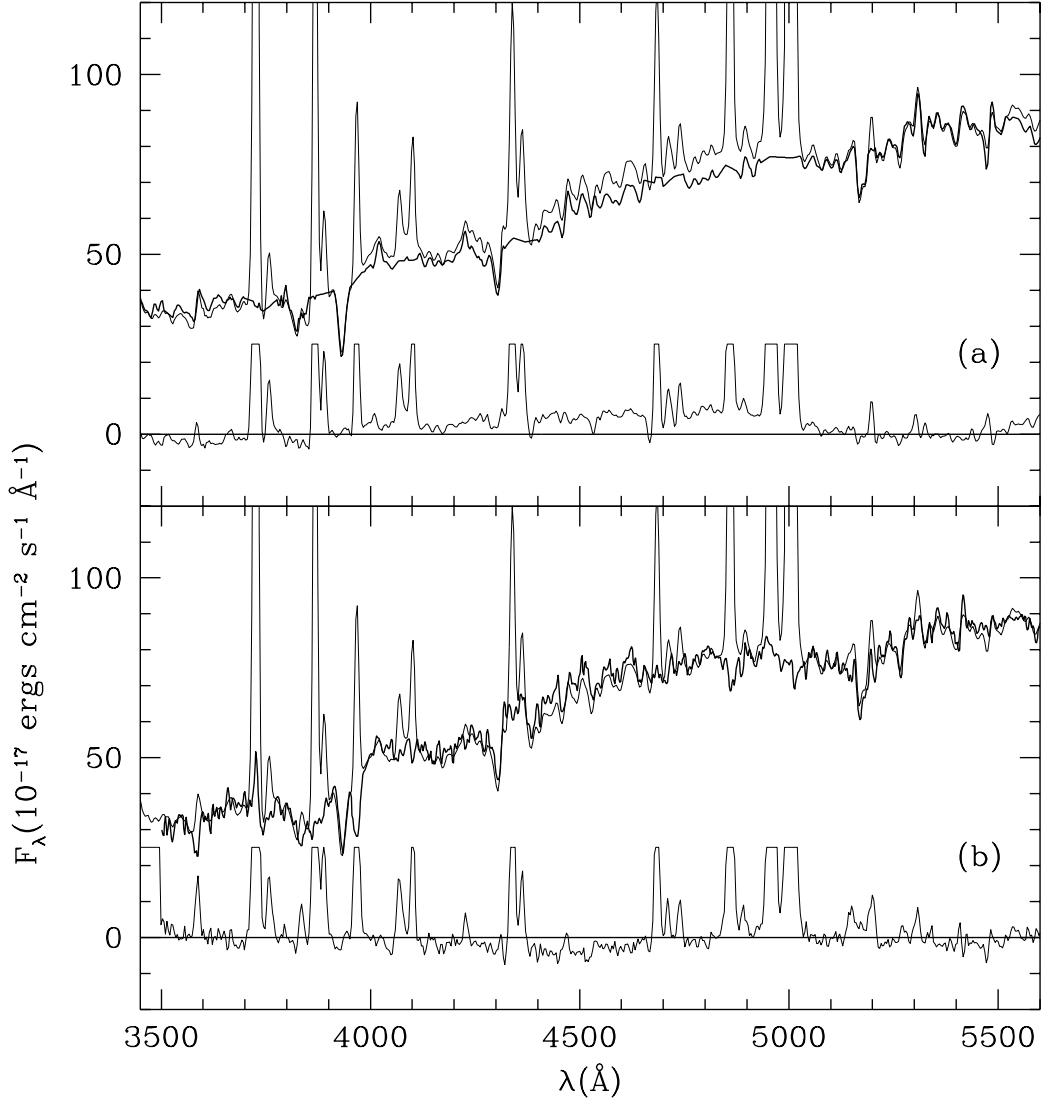}
\caption{ Comparison between the nuclear and extranuclear bulge
spectra of Mrk\,573: (a) The nuclear spectrum of Mrk\,573 (thin line) is 
compared with the normalized and reddened [E(B-V)=0.10] 
extranuclear spectrum (thick line), where the emission lines have been chopped for clarity. The residual between the two is plotted at the bottom of the 
figure. (b) The nuclear spectrum (thin line) compared with the
elliptical template (IC\,4889, thick line) combined with
a $F_\lambda \propto \lambda^{-1}$  power-law, with contributions of 90\% and 10\%, respectively, to the total flux at $\lambda$5500\AA. The residual
between the two is plotted at the bottom, showing a better agreement
for case (b). The emission lines
in the residual spectra have also been chopped for clarity.}
\end{figure}

\begin{figure}
\plotone{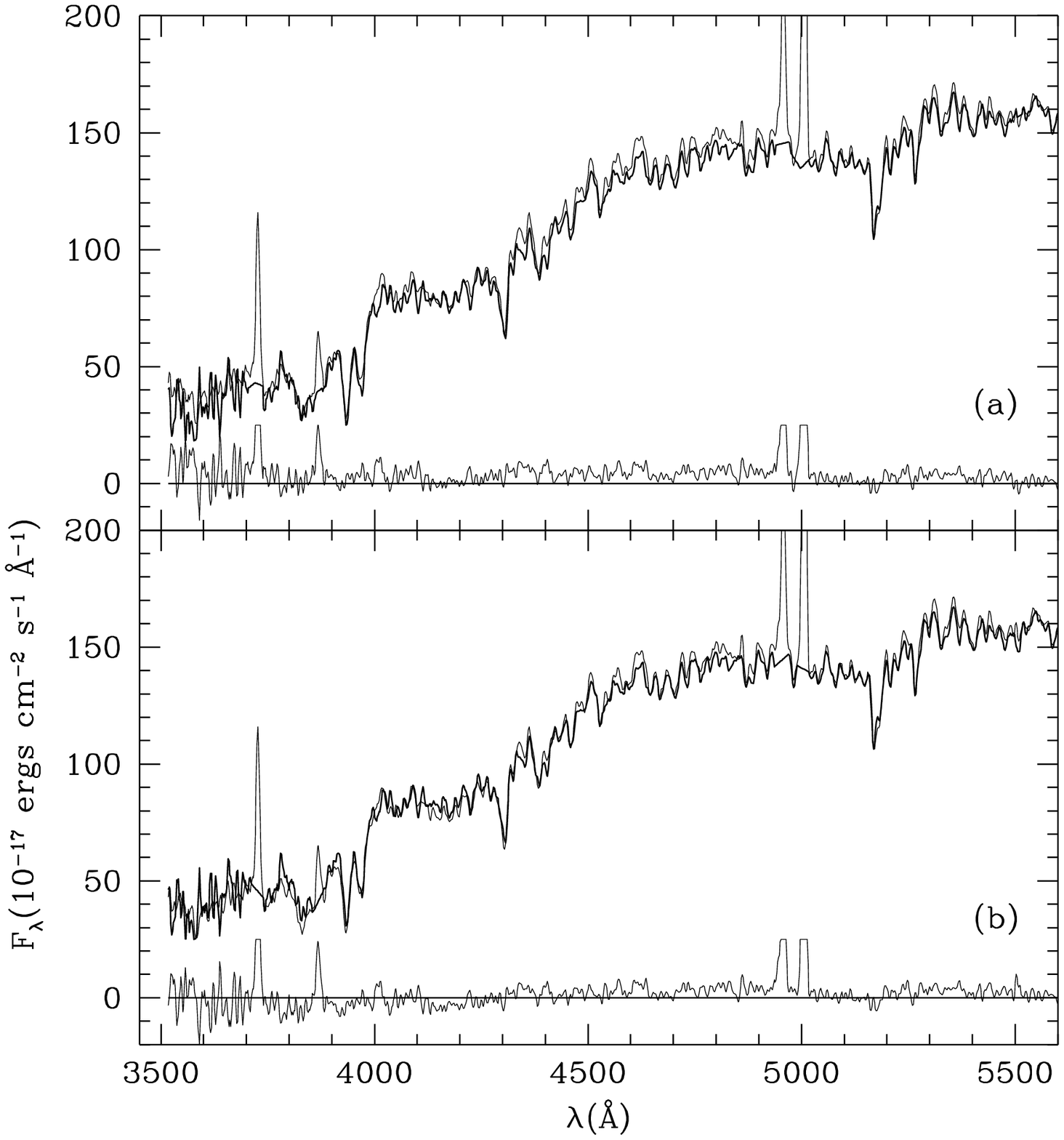}
\caption{(a) The nuclear spectrum of NGC\,1358 (thin line) compared with 
the normalized and reddened [E(B-V)=0.10]  
extranuclear spectrum (thick line), 
where the emission lines have been chopped for clarity. 
The residual between the two is plotted at the bottom of the figure. 
(b) The nuclear spectrum (thin line) is compared with the
extranuclear spectrum reddened by E(B-V)=0.10 and combined with
a 3\% contribution (at $\lambda$5500\AA) of the FC1 continuum, 
represented by a $F_\lambda \propto \lambda^{-1}$ power-law (thick line). 
The residual between the two is plotted at the bottom. The emission lines
in the residual spectra have also been chopped for clarity.}
\end{figure}

\begin{figure}
\plotone{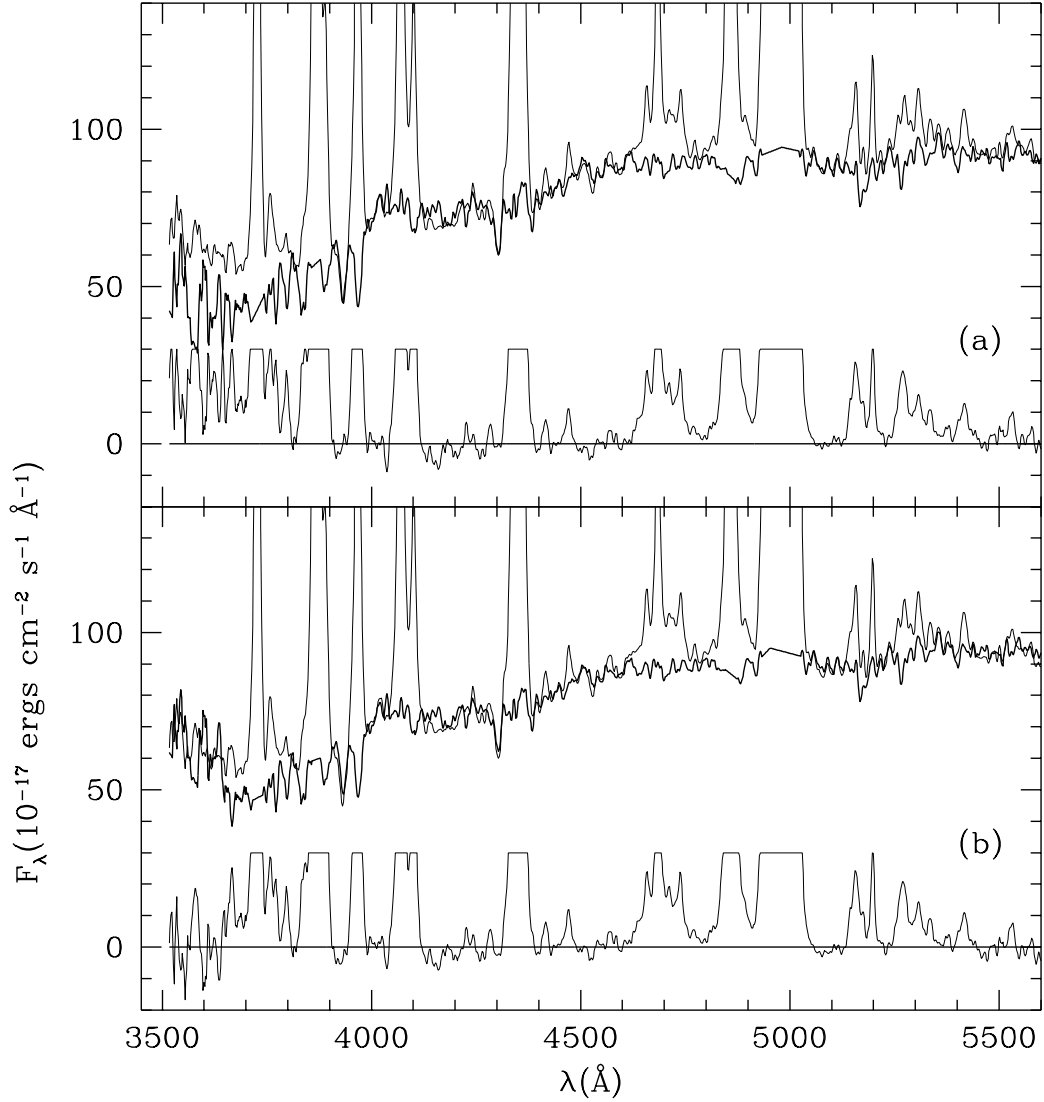}
\caption{(a) The nuclear spectrum of Mrk\,1210 (thin line) compared with
the normalized and reddened [E(B-V)=0.07]
 extranuclear spectrum (thick line), where the emission lines have been chopped for clarity. The residual between the two is plotted at the bottom of the 
figure. (b) The nuclear spectrum (thin line) compared with the
extranuclear spectrum reddened by [E(B-V)=0.15],
combined with the Balmer continuum and
a 6\% contribution of a FC1 component $F_\lambda \propto \lambda^{-0.7}$.
The residual between the two is plotted at the bottom. Notice the better
agreement with the new components. The emission lines
in the residual spectra have also been chopped for clarity.}
\end{figure}

\begin{figure}
\plotone{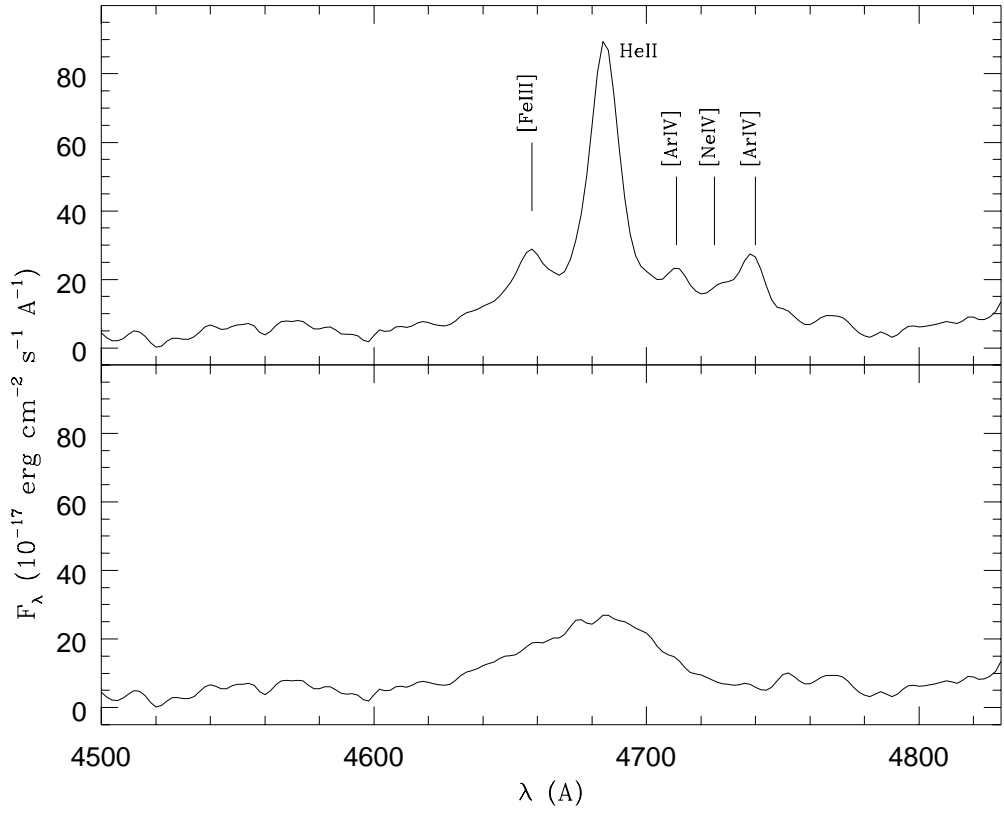}
\caption{(a) The spectrum of Mrk\,1210 in the region of He II$\lambda$4686\AA;
(b) the WR feature: residual after the subtraction of the narrow emission lines
Fe[III]$\lambda$4658\AA, HeII$\lambda$4686\AA,
[ArIV]$\lambda$4711,40\AA\ and [NeIV]$\lambda$4725\AA.}
\end{figure}

\begin{figure}
\plotone{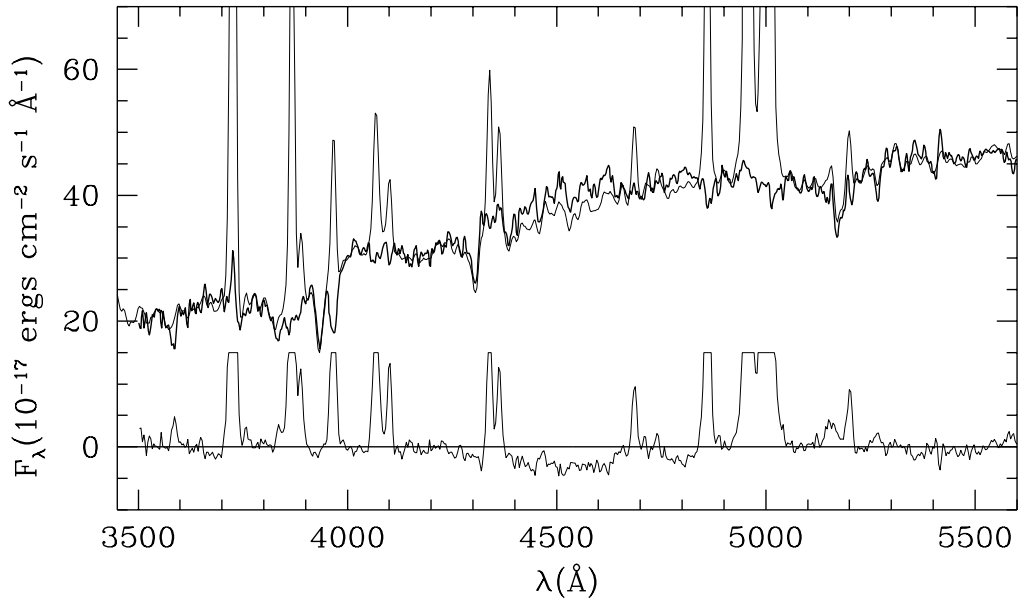}
\caption{Top: Comparison of the nuclear spectrum (thin line) 
of Mrk\,348 with the elliptical template (thick line) 
combined with a 15\% contribution
(at $\lambda$5500\AA) of a $\lambda^{-1}$ power-law component. Bottom:
residual between the two, where the emission lines
have been chopped for clarity.}
\end{figure}

\begin{figure}
\plotone{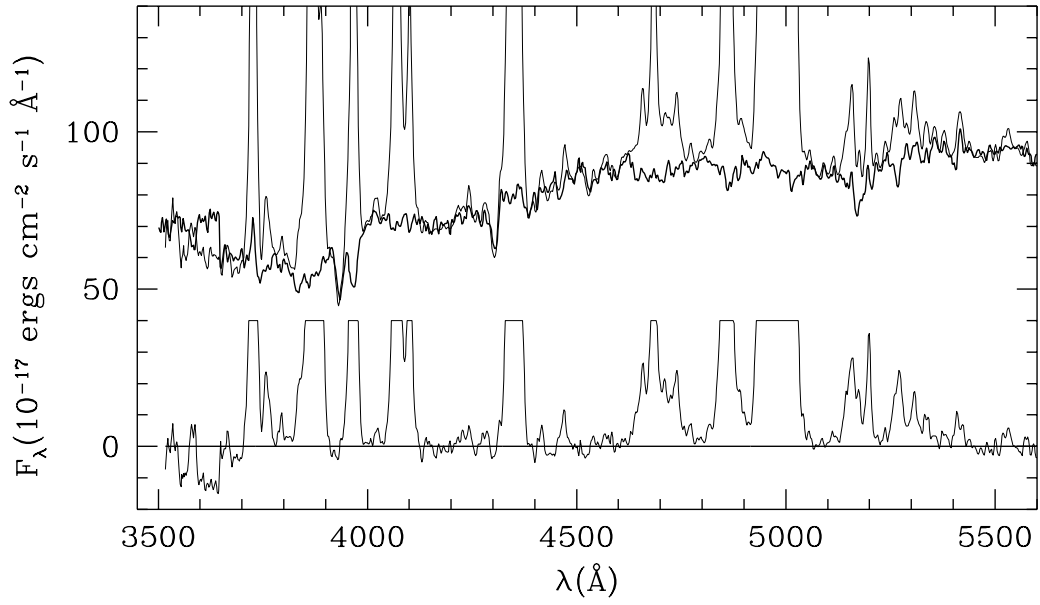}
\caption{Top: Comparison of the nuclear spectrum (thin line) 
of Mrk\,1210, with the
elliptical template  (thick line) combined with a 28\% contribution
(at $\lambda$5500\AA) of a $\lambda^{-0.7}$ power-law component 
and 3\% of the Balmer continuum. Bottom: residual
between the two, where the emission lines
have been chopped for clarity.}
\end{figure}

%
%
\clearpage


\begin{deluxetable}{lcccc}
\tablewidth{0pc}
\tablecaption{Log of Observations}
\tablehead{\colhead{Galaxy}&\colhead{Date}&\colhead{Exp. Time (sec)}&\colhead{Pos. Angle($^\circ$)}&\colhead{E(B-V)$_G$}}
\startdata
Mrk\,348       &6/7 Dez. 94 &1800         &170         &0.060\cr
Mrk\,573       &6/7 Dez. 94 &1800         &125         &0.008\cr
NGC\,1358      &6/7 Jan. 94 &900          &145         &0.025\cr
Mrk\,1210      &5/6 Jan. 94 &1800         &163         &0.018\cr
IC\,4889       &28/29 May 92 &1800        &0           &0.045\cr
\enddata
\end{deluxetable}


\begin{deluxetable}{lccccc}
\tablewidth{0pc}
\tablecaption{Population synthesis results\tablenotemark{1}}
\tablehead{\colhead{Pop. template}&\colhead{10 Gyr}&\colhead{1 Gyr}&\colhead{100 Myr}&\colhead{10 Myr}&\colhead{HII}}
\startdata
Mrk\,348       & 45 (36)  & 45 (50)  & 5 (8)    & 4 (5)   & 1 (1)\cr
NGC\,1358      & 64 (55)  & 34 (42)  & 2 (3)    & 0 (0)   & 0 (0)\cr
Mrk\,1210      & 53 (42)  & 24 (25)  & 21 (31)  & 2 (2)   & 0 (0)\cr
IC\,4889       & 79 (72)  & 21 (28)   & 0 (0)    & 0 (0)   & 0 (0)\cr
\tablenotetext{1}{Percent contribution of different age bins to the
light at $\lambda$5870\AA\  (and $\lambda$4510\AA)}
\enddata
\end{deluxetable}

\end{document}